\documentclass[english]{article}
\usepackage[latin9]{inputenc}
\usepackage{graphicx}
\usepackage{hyperref}
\usepackage{multirow}
\usepackage{authblk}
\usepackage{slashed}
\usepackage{cite}
\usepackage[top=3.5cm]{geometry}

\makeatletter

\providecommand{\tabularnewline}{\\}

\makeatother

\usepackage{babel}
\usepackage{amsmath}

\title{The Lepton-Gluon Portal Beyond Lepto-Gluons}
\author[1]{Linda M. Carpenter\thanks{lmc@physics.osu.edu}}
\author[1]{Katherine Schwind\thanks{schwind.44@osu.edu}}

\affil[1]{\emph{Department of Physics, The Ohio State University,
191 W. Woodruff Avenue, Columbus, OH 43210, U.S.A.}}

\begin{document}

\maketitle

\begin{abstract}

We explore models where single new exotic states interact with the
Standard Model through an asymmetric Standard Model portal with couplings
to at least one gluon and one lepton. We consider the complete set of effective operators up to dimension 6, and examine a few additional dimension 7 operators that contain interesting field content or potential collider signals. The lepton-gluon portal allows access to exotic states with an interesting range of SU(3) and SU(2) quantum numbers. Finally, we explore potential single-production modes and their phenomenological signatures at colliders.
\end{abstract}

\newpage

\section{Introduction}
This work is part of a series which attempts to make a general map of the Light Exotic (LEX) Beyond the Standard Model (BSM) particles which can be discovered through specific collider processes. These works have been part of a program called Light Exotics Effective Field Theory (LEX-EFT) \cite{Carpenter:2023giu}, where we make a first pass at categorizing the interactions of BSM particles by capturing their interactions with the Standard Model (SM) through a series of effective operators. There are many directions which effective field theory can take, but for the purpose of this paper we are using what we term a ``portal-based" approach to mapping BSM theories. In this approach, we choose a fixed set of SM fields and then determine the quantum numbers of new particles that can be discovered through interactions with this ``portal" \cite{Carpenter:2024hvp,Carpenter:2024jje,Carpenter:2025cys,Carpenter:2021gpl}. 

The portal-based approach outlines which species of BSM particles can be accessed through specific collider processes. This exercise does not discriminate between theories, but rather keeps consideration open to all manner of BSM particle types. Only a subsection of potential BSM parameter space has been previously explored in various phenomenological paradigms---in general, the literature is focused on new states in somewhat low representations of  SU(3) and SU(2). However, the portal-based approach opens up more of the possible theory space to phenomenological consideration, and provides a more agnostic framework from which to study collider phenomenology.  

In this work, we will study a lesser-explored portal to new physics, the lepton-gluon ($\ell$-G) portal. That is, here we will consider effective field theory interactions in which new LEX states interact with the  SM in interactions that involve at least one lepton and one gluon. These interactions may, or may not, contain additional BSM states. We can visualize the resultant
interactions in Fig. 1, where $X$ is a LEX state.
\begin{center}
    \includegraphics[width=0.6\linewidth]{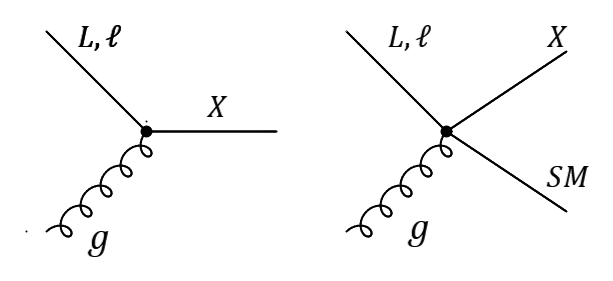}\\
    {Figure 1. Possible interaction vertices in the lepton-gluon portal}
    \label{fig:placeholder}\\
    \end{center}
Such interactions will allow us to study the production and decay of a large number of exotic particles that carry both SU(3) and electroweak quantum numbers, and that carry non-zero lepton number. 

We note that previous works have studied the lepto-gluon, a possible exotic fermionic octet which couples only to one gluon and one lepton \cite{Han:2025wdy,Goncalves-Netto:2013nla,Sahin:2010dd,Jelinski:2015epa,Mandal:2016csb,Streng:1986my}. The lepto-gluon is an example of an exotic color-charged state which (like leptoquarks) carries lepton number. Here, we will see that the lepton-gluon portal allows us to access even more exotic particles. The general operator analysis presents windows to very interesting regions of theory space, some of which have scarcely been studied. These regions include theories of states in exotic color representation such as 15's of SU(3) ---in addition to the  color sextets--- bi-adjoints of the SU(3) and SU(2) SM gauge groups, and fermionic color octets (species of lepto-gluon) with higher SU(2) charge. We also find interesting new effective interactions for some more  well-known BSM particles such as scalar leptoquarks \cite{BUCHMULLER1987442} and singlet color octets \cite{Chen:2014haa,Gerbush:2007fe,Plehn:2008ae}.

In this work, we concentrate on writing operators in which a single LEX state may be produced or decay. The reason for this is several-fold. First, these types of operators will allow for the decay of exotic states into SM particles. Second, while states of any SM charge can always be pair produced through interactions with gauge bosons, the operators in this work allow for the study of single production of exotic particles in association with one or more SM particles. It has been shown that in many instances, single production induced by an effective operator is the dominant collider production mode for LEX states, overtaking pair production when particle masses begin to exceed the 1 TeV range. 

We will catalog all possible effective operators, up to dimension 6, in which a single exotic particle interacts with the SM through interactions that contain at least one lepton and one gluon. For each of these operators, we will specify the quantum numbers of the exotic state included. While we will not make a complete list of operators at dimension 7, we will note the most interesting operators in the lepton-gluon portal at this dimension. In constructing the lepton-gluon portal, we explain how we choose a phenomenological operator basis which maximizes the number of operators. 

This paper proceeds as follows. In Section 2, we will enumerate all possible operators, up to dimension 6, which contain a scalar (spin-0) BSM state. In Section 3, we similarly discuss operators which contain fermionic (spin-1/2) LEX states. Section 4 briefly investigates a few dimension 7 operators. Section 5 focuses on some phenomenological signatures that operators in this portal could cause. Section 6 concludes.

\section{Spin 0 LEX States}
We start our catalog by enumerating all operators, up to dimension 6, which couple BSM states to at least one SM lepton and one gluon. All operators must be a singlet under all SM gauge groups. As such, we can iteratively find all SM quantum numbers which the LEX state may carry in each case of set SM fields. In the tables below, we write the operators themselves, the allowed SM SU(3), SU(2), and U(1)$_Y$ quantum numbers of the LEX state, and the baryon number (B) and lepton number (L) of the LEX state. We acknowledge that in some theories, baryon number and lepton number may be violated; however, we write these for the LEX state in the assumption that these are not violated. In the operator tables below, Lorentz indices are written with Greek indices, while SU(2) fundamental indices are written using lowercase Roman indices. We choose not to explicitly write out our SU(3) indices.

In this section, we will specifically write operators which contain a CP-even, spin-0 (scalar) LEX state. We know that all operators in the lepton-gluon portal must contain at least one lepton, which may be left- or right-handed. We also know that operators must contain an even number of fermions; as our LEX state is a scalar, operators in this section require an additional SM fermion, which can be either a quark or a lepton. We can generate our desired coupling to a gluon through either the insertion of an SU(3) field strength tensor or the application of a covariant derivative to a state which has non-singlet SU(3) charge.

\subsection{$\boldsymbol{GLf\phi}$}
There is only one operator type in the lepton-gluon portal that involves a scalar LEX field and no derivatives. By necessity, these operators contain the scalar LEX field, one lepton, one other SM fermion (either a quark or a second lepton), and the SU(3) field tensor, $G_{\mu\nu}$. A schematic of the 4-particle vertices induced by these operators is shown in Fig. 2. 
\begin{center}
    \includegraphics[width=0.3\linewidth]{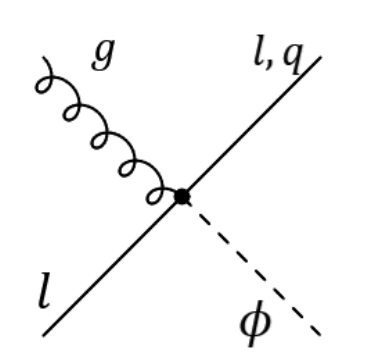}\\
    {Figure 2. Four particle vertices of type $Glf\phi$}
    \label{fig:placeholder}\\
    \end{center}
In these operators, both of the SM fermions can, independently, be either left- or right-handed. These operators, presented in Table I, are dimension 6.

In order to find the quantum numbers of the LEX state, we follow known tensor product rules. To determine the LEX SU(3) numbers, we look to the quark and gluon content of each operator. In the case that the operator contains both an SU(3) field strength tensor and a quark, we look to the SU(3) product rule $8\otimes 3 = 15 \oplus\overline{6} \oplus 3$ \cite{Slansky:1981yr}. As such, the LEX state can be a color 15-plet, sextet, or triplet. For operators with an SU(3) field strength tensor and no quarks, the LEX state must be a color octet. 

Next, we determine the SU(2) numbers of our LEX state. In the operators of this section, the only SM fields which may carry SU(2) charge are the two fermions. In the case where both SM fermions are right-handed, all SM fields within our operator are SU(2) singlets. As such, the LEX state will also be an SU(2) singlet. When  one of the SM fermions is right-handed and the other left-handed, the LEX state must be an SU(2) doublet. When both SM fermions are left-handed, we follow the SU(2) product rule $2 \otimes 2 = 1 \oplus 3$; the LEX state can be either an SU(2) singlet or triplet.

\begin{center}
\begin{tabular}{|c|c|c|c|}
\hline 
Operators with scalar LEX fields & (SU(3),SU(2),Y) & B & L \tabularnewline
\hline 
\hline 
\multirow{2}{*}{$G_{\mu\nu} \overline{L^c}_i \sigma^{\mu \nu} Q_{L j} \phi^{ij}$} & ($\overline{15},3,1/3$), ($6,3,1/3$), & $-1/3$ & $-1$ \tabularnewline
& ($\overline{3},3,1/3$) & & \tabularnewline
\hline 

\multirow{2}{*}{$G_{\mu\nu} \overline{\ell} \sigma^{\mu \nu} Q_{L i} \phi^i$} & ($\overline{15},2,-7/6$), ($6,2,-7/6$), & \multirow{2}{*}{$-1/3$} & \multirow{2}{*}{$1$} \tabularnewline
& ($\overline{3},2,-7/6$) & & \tabularnewline
\hline 
\multirow{2}{*}{$G_{\mu\nu} \overline{L}^i \sigma^{\mu \nu} u \phi_i$} & ($\overline{15},2,-7/6$), ($6,2,-7/6$), & \multirow{2}{*}{$-1/3$} & \multirow{2}{*}{1} \tabularnewline
& ($\overline{3},2,-7/6$) & & \tabularnewline
\hline 
\multirow{2}{*}{$G_{\mu\nu} \overline{L}^i \sigma^{\mu \nu} d \phi_i$} & ($\overline{15},2,-1/6$), ($6,2,-1/6$), & \multirow{2}{*}{$-1/3$} & \multirow{2}{*}{1} \tabularnewline
& ($\overline{3},2,-1/6$) & & \tabularnewline
\hline 

\multirow{2}{*}{$G_{\mu\nu} \overline{\ell^c} \sigma^{\mu \nu} u \phi$} & ($\overline{15},1,1/3$), ($6,1,1/3$), & \multirow{2}{*}{$-1/3$} & \multirow{2}{*}{$-1$} \tabularnewline
& ($\overline{3},1,1/3$) & & \tabularnewline
\hline 
\multirow{2}{*}{$G_{\mu\nu} \overline{\ell^c} \sigma^{\mu \nu} d \phi$} & ($\overline{15},1,4/3$), ($6,1,4/3$), & \multirow{2}{*}{$-1/3$} & \multirow{2}{*}{$-1$} \tabularnewline
& ($\overline{3},1,4/3$) & & \tabularnewline
\hline 
\multirow{2}{*}{$G_{\mu\nu} \overline{L^c}^i \sigma^{\mu \nu} Q_{L i} \phi$} & ($\overline{15},1,1/3$), ($6,1,1/3$), & $-1/3$ & $-1$ \tabularnewline
& ($\overline{3},1,1/3$) & & \tabularnewline
\hline 
\hline
$G_{\mu\nu} \overline{L^c}_i \sigma^{\mu \nu} L_j \phi^{ij}$ & ($8,3,1$) & $0$ & $-2$ \tabularnewline
\hline 
$G_{\mu\nu} \overline{\ell} \sigma^{\mu \nu} L_i \phi^i$ & ($8,2,-1/2$) & 0 & 0 \tabularnewline
\hline 
$G_{\mu\nu} \overline{L^c}^i \sigma^{\mu \nu} L_i \phi$ & ($8,1,1$) & $0$ & $-2$ \tabularnewline
\hline
$G_{\mu\nu} \overline{\ell^c} \sigma^{\mu \nu} \ell \phi$ & ($8,1,2$) & $0$ & $-2$\tabularnewline
\hline

\end{tabular}
\\
Table I. Operators containing $GLf\phi$
\end{center}

Due to chirality projectors, operators that contain two right-handed or two left-handed fermions must contain a fermion and fermion-bar conjugate pair. Operators with one left-handed and one right-handed fermion must contain a fermion anti-fermion pair. 

LEX states in operators with one fermion and one quark (the 15-plet, sextet, and triplet) may be assigned leptoquark-type baryon and lepton numbers. In fact, the LEX states in the fundamental SU(3) representation are types of leptoquarks. The exotic sextets we will refer to as lepto-di-quarks, while the 15-plets are dekapenta-leptoquarks. 
For brevity, we note that LEX states in 15, $\overline{6}$, and $3$ representations may be found by barring the fermion bi-linears in the table. Most impressively, we see that the highest SM representation we can access in the lepton-gluon portal is the ($\overline{15},3,1/3$) state. Work on the 15-plet representations is very thin on the ground. Sextet fermions are another lesser explored SU(3) representation. There have been dedicated collider studies to renormalizable sextet couplings couplings\cite{Fortes:2013dba,Arina:2025zpi,Han:2010rf,Han:2023djl}. However, EFT operators have been less studied. The authors have explored some EFT couplings of lepto-diquark sextets in some lower SU(2) representations \cite{Carpenter:2021rkl,Carpenter:2023aec,Carpenter:2022qsw}, but LHC searches in the full parameter space  involving higher SU(2) representations have yet to be completed.

The operators containing LEX scalars that are octets of SU(3) have baryon number 0 and lepton number
$-2$ or 0.  The unusual states with double lepton number come from operators that must contain a lepton and charge conjugated anti-lepton. The scalar octet with lepton number 0 comes from an operator that contains a gluon field strength tensor, LEX scalar, lepton, and anti-lepton. Interestingly, this state must be an SU(2) doublet and has identical quantum numbers to the Manohar-Wise field\cite{Manohar:2006ga}.

The ($8,2,-1/2$) particle has charge components ($\phi_0,\phi_{-}$). Expanding its couplings in the lepton-gluon portal, we obtain
\begin{equation}
G_{\mu\nu}\overline{\ell}\sigma^{\mu\nu}\nu\phi_{-} \quad \mathrm{and} \quad G_{\mu\nu}\overline{\ell}\sigma^{\mu\nu}\ell\phi_{0} \ .
\end{equation}
The above couplings are new dimension 6 interactions for this field.

The scalar octets with double lepton number have not (as of yet) appeared in the literature. The SU(2) triplet ($8,3,1$) multiplet contains three charged components, as shown in Eq. 2.
\begin{equation}
\begin{pmatrix}
\phi_{-}/\sqrt{2} & \phi_0  \\
\phi_{--}  & \phi_{-}/\sqrt{2} 
\end{pmatrix}  
\end{equation}
The differently charged components each couple to two leptons, two neutrinos, or a single lepton and single neutrino. The ($8,3,1$) field is a scalar bi-adjoint of the strong and weak gauge groups. The authors have recently begun explorations into the parameter space of simple scalar bi-adjoints \cite{Carpenter:2024qti,Carpenter:2022oyg}, but this representation is phenomenologically underexplored.

The ($8,1,-1$) field is a singly charged color octet scalar, while the ($8,1,2$) field is a doubly charged color octet scalar. This doubly charged field couples simply to a charged lepton pair: $G_{\mu\nu}\ell^{\dagger C}\sigma^{\mu\nu}\ell\phi_{++}$.

We note that we may split the masses of the components of the triplet and doublet octet LEX states through renormalizable interactions with the Higgs field. Writing out SU(2) indices explicitly for color octet states that are either SU(2) doublets or triplets, we may have
\begin{equation}
L \supset \lambda_2(\phi^{\dagger i}H_i)(\phi_jH^{\dagger j}) 
\quad \mathrm{or} \quad
 \lambda_3(\phi^{\dagger ij}H_i)(\phi_{jk}H^{\dagger k}) \ .
\end{equation}
We may thus induce cascade decays among the charge components of the color octet triplet or doublet. These couplings will affect the overall loop-induced Higgs couplings to gluons and will be constrained by Higgs production and decay measurements. However, it is reasonable to believe that octets in the TeV range with Higgs couplings in the range of 0.1 would be unconstrained \cite{Boughezal:2010kx, Cao:2013wqa}. This would allow splittings of order 10-50 GeV.

\subsubsection*{Other Lorentz structures for $\boldsymbol{GLf \phi}$}
The operators in Table I represent the simplest Lorentz structures that contain the fermion tensor current $\overline{f}\sigma^{\mu\nu}f$ and a CP even spin-0 scalar $\phi$.  There are several other possible Lorentz structures that involve both the aforementioned scalar $\phi$ and CP odd spin 0 pseudo-scalar $\tilde{\phi}$. Pseudo-scalar production is generally known to be larger than scalar production at hadron colliders \cite{Field:2002gt}, so these operators may be important discovery channels. The operators with alternate Lorentz structures are
\begin{equation}
\tilde{G}_{\mu\nu}\overline{f}\sigma^{\mu\nu}f\tilde{\phi} \ . 
\end{equation}
The operators in Table I may be altered to conform to either of this Lorentz structure while maintaining the same SM charge assignments for the LEX states.

\subsection{$\boldsymbol{DLQ\phi}$}

Next, we consider operators which do not contain an SU(3) field strength tensor. Instead, we can generate an off-shell gluon through use of a derivative acting on a field with non-singlet SU(3) charge. Specifically, we know that 
\begin{equation}
    D^{\mu}\Phi \supset \partial^{\mu} \Phi + i g_3 \tau_3  A_3^{\mu}\Phi +i g_2 \tau_2 A_2^{\mu}\Phi  + ig_1 Y_{\phi} A_1^{\mu}\Phi \ ,
\end{equation}
where $\Phi$ is a field charged under each of the (SU(3), SU(2), SU(1)) groups. Here, the gluon is included in the $A_3$ gauge components.

We begin by considering operators with a lepton, a second SM fermion, a LEX scalar, and one derivative. These operators will be dimension 5. In order to introduce SU(3) charge and allow the generation of a gluon, we must choose the second SM fermion to be a quark. In the following computations, we will use equations of motion, integration by parts, and other field theory relations to remove redundant operators in the portal. These techniques are similar to those in SMEFT theories \cite{Grzadkowski_2010,Murphy:2020rsh,Lehman:2014jma}. However, we will not insist on using an on-shell basis-similar to the Warsaw basis of SMEFT. Instead, we will choose a phenomenological basis that keeps all non-redundant operators in order to maximize the utility of the lepton-gluon portal. 

At first glance, we can choose the derivative  of the $DLQ\phi$-type operator to act on any of the three fields in this operator. However, operators with the derivative acting on the lepton do not contribute to the $\ell$-G portal. We then have two operators left in the portal and one integration by parts relation:
\begin{equation}
 D_\mu \overline{L} \gamma^{\mu}  Q \phi +   \overline{L} \gamma^{\mu} D_\mu Q \phi = -\overline{L} \gamma^{\mu}  Q D_\mu \phi \ .
\end{equation}
We choose to discard the operators where the derivative acts on the LEX scalar. We note that, as the remaining operator is proportional to a fermion equation of motion, it could be removed in an on-shell basis similar to the Warsaw basis of SMEFT. However, since we have a choice of basis, we choose a maximalist interpretation of the lepton-gluon portal which contains the operator. Operators of this type are shown in Table II.
\begin{center}
\begin{tabular}{|c|c|c|c|}
\hline 
Operators with scalar LEX fields & (SU(3),SU(2),Y) & B & L \tabularnewline
\hline 
\hline 
$ \overline{\ell} \gamma^{\mu} D_\mu u \phi$ & ($\overline{3},1,-5/3$) & $-1/3$ & $1$ \tabularnewline
\hline 
$ \overline{\ell} \gamma^{\mu}D_\mu d \phi$ & ($\overline{3},1,-2/3$) & $-1/3$ & $1$ \tabularnewline
\hline 
$ \overline{L^c}_i \gamma^{\mu} D_\mu u \phi^i$ & ($\overline{3},2,-1/6$) & $-1/3$ & $-1$ \tabularnewline
\hline 
$ \overline{L^c}_i \gamma^{\mu}D_\mu d \phi^i$ & ($\overline{3},2,5/6$) & $-1/3$ & $-1$ \tabularnewline
\hline
$ \overline{\ell^c} \gamma^{\mu} D_\mu Q_{L i} \phi^i$ & ($\overline{3},2,5/6$) & $-1/3$ & $-1$ \tabularnewline
\hline 
$ \overline{L}^i \gamma^{\mu} D_\mu Q_{L j} \phi_i^{j}$ & ($\overline{3},3,-2/3$) & $-1/3$ & $1$ \tabularnewline
\hline 
$ \overline{L}^i \gamma^{\mu} D_\mu Q_{L i} \phi$ & ($\overline{3},1,-2/3$) & $-1/3$ & $1$ \tabularnewline
\hline 

\end{tabular}
\\
Table II. Operators containing $DLQ\phi$
\end{center}

In Table II,  operators with two right-handed SM fermions result in a LEX state that is an SU(2) singlet, while those with one left-handed and one right-handed fermion lead to a LEX SU(2) doublet. Furthermore, operators with two left-handed SM fermions again result in either singlet or triplet LEX states. The SU(3) numbers of the LEX states in Table II are straightforward; as all of the operators in this table contain only a single quark as their SM field content with SU(3) charge, the LEX field must be a color anti-triplet. All LEX states are assigned baryon number $-1/3$ and lepton number 1. These states are therefore species of scalar leptoquarks.

\subsection{$\boldsymbol{DLQ\phi H}$}
Next, we consider operators with two SM fermions, a LEX scalar, one derivative, and a Higgs field. These operators are dimension 6. Again, we require one SM fermion to be a lepton, in order to satisfy the $\ell$-G portal requirement, and the other to be a quark, in order to generate SU(3) charge. The derivative may once again act on either the quark or the LEX state, with one integration by parts relation relating these operators. We again discard the choice where the derivative acts on the LEX field. These operators can be found in Table III.
\begin{center}
\begin{tabular}{|c|c|c|c|}
\hline 
Operators with scalar LEX fields & (SU(3),SU(2),Y)& B & L\tabularnewline
\hline 
\hline 
 $ \overline{\ell} \gamma^{\mu} D_\mu u \phi^i H_i$ & ($\overline{3},2,-13/6$) & $-1/3$ & $1$ \tabularnewline
\hline 
 $ \overline{\ell} \gamma^{\mu}D_\mu d \phi^i H_i$ & ($\overline{3},2,-7/6$) & $-1/3$ & $1$ \tabularnewline
\hline 
 $ \overline{L^c}_i \gamma^{\mu} D_\mu u \phi^{ij} H_j$ & ($\overline{3},3,-2/3$) & $-1/3$ & $-1$ \tabularnewline
\hline
 $ \overline{L^c}_i \gamma^{\mu} D_\mu d \phi^{ij} H_j$ & ($\overline{3},3,1/3$) & $-1/3$ & $-1$ \tabularnewline
\hline
 $ \overline{L^c}^i \gamma^{\mu} D_\mu u \phi H_i$ & ($\overline{3},1,-2/3$) & $-1/3$ & $-1$ \tabularnewline
\hline
 $ \overline{L^c}^i \gamma^{\mu} D_\mu d \phi H_i$ & ($\overline{3},1,1/3$) & $-1/3$ & $-1$ \tabularnewline
\hline
 $ \overline{\ell^c} \gamma^{\mu} D_\mu Q_{L i} \phi^{ij}H_j$ & ($\overline{3},3,1/3$) & $-1/3$ & $-1$ \tabularnewline
\hline 
 $ \overline{\ell^c} \gamma^{\mu} D_\mu Q_{L }^i \phi H_i$ & ($\overline{3},1,1/3$) & $-1/3$ & $-1$ \tabularnewline
\hline 
 $ \overline{L}^i \gamma^{\mu} D_\mu Q_{L j} \phi_i^{jk} H_k$ & ($\overline{3},4,-7/6$) & $-1/3$ & $1$ \tabularnewline
\hline 
 $ \overline{L}^i \gamma^{\mu} D_\mu Q_{L i} \phi^j H_j$, \quad $\overline{L}^i \gamma^{\mu} D_\mu Q_{L j} \phi^j H_i$ & ($\overline{3},2,-7/6$) & $-1/3$ & $1$ \tabularnewline
\hline 

 $ \overline{\ell} \gamma^{\mu} D_\mu u \phi_i H^{\dagger i}$ & ($\overline{3},2,-7/6$) & $-1/3$ & $1$ \tabularnewline
\hline 
 $ \overline{\ell} \gamma^{\mu}D_\mu d \phi_i H^{\dagger i}$ & ($\overline{3},2,-1/6$) & $-1/3$ & $1$ \tabularnewline
\hline 
 $ \overline{L^c}_i \gamma^{\mu} D_\mu u \phi^{i}_j H^{\dagger j}$ & ($\overline{3},3,1/3$) & $-1/3$ & $-1$ \tabularnewline
\hline
 $ \overline{L^c}_i \gamma^{\mu} D_\mu d \phi^{i}_j H^{\dagger j}$ & ($\overline{3},3,4/3$) & $-1/3$ & $-1$ \tabularnewline
\hline
 $ \overline{L^c}_i \gamma^{\mu} D_\mu u \phi H^{\dagger i}$ & ($\overline{3},1,1/3$) & $-1/3$ & $-1$ \tabularnewline
\hline
 $ \overline{L^c}_i \gamma^{\mu} D_\mu d \phi H^{\dagger i}$ & ($\overline{3},1,4/3$) & $-1/3$ & $-1$ \tabularnewline
\hline
 $ \overline{\ell^c} \gamma^{\mu} D_\mu Q_{L i} \phi^{i}_j H^{\dagger j}$ & ($\overline{3},3,4/3$) & $-1/3$ & $-1$ \tabularnewline
\hline 
 $ \overline{\ell^c} \gamma^{\mu} D_\mu Q_{L i } \phi H^{\dagger i}$ & ($\overline{3},1,4/3$) & $-1/3$ & $-1$ \tabularnewline
\hline 
 $ \overline{L}^i \gamma^{\mu} D_\mu Q_{L j} \phi_{ik}^{j} H^{\dagger k}$ & ($\overline{3},4,-1/6$) & $-1/3$ & $1$ \tabularnewline
\hline 
 $ \overline{L}^i \gamma^{\mu} D_\mu Q_{L i} \phi_j H^{\dagger j}$, \quad $\overline{L}^i \gamma^{\mu} D_\mu Q_{L j} \phi^j H^\dagger_i$ & ($\overline{3},2,-1/6$) & $-1/3$ & $1$ \tabularnewline
\hline 

\end{tabular}
\\
Table III. Operators containing $DLQ\phi H$
\end{center}

In the operators of Table III, there are three SM fields which may carry non-singlet SU(2) charge. In the case where both of the quarks are left-handed, all three of these SM fields are SU(2) doublets. As such, we can look to the SU(2) product rule
\begin{equation}
    2\otimes2\otimes2 = 2\otimes (1\oplus3) = 2\oplus4 \ ;
\end{equation}
the LEX state may be an SU(2) doublet or quadruplet. The case where one SM fermion is left-handed and one is right-handed contains a total of two SM fields that are SU(2) doublets. As previously discussed in this work, this leads to LEX states that are either SU(2) singlets or triplets. Finally, when both SM fermions are right-handed, the only SM field with non-singlet SU(2) charge is the Higgs; as such, the LEX state can only be an SU(2) doublet. Once again, the only SM field charged under SU(3) is the quark. The LEX state again has anti-triplet color charge. We also note that Higgs vevs may be inserted into any operator to produce operators of effective dimension 5 with effective cutoff $v_h/\Lambda^2$.

The accessible states of this type all carry baryon number $-1/3$ and unit lepton number. These states are all types of scalar leptoquarks. There has been some previous EFT work focused on dimension 5 and 6 leptoquark operators coupling to Higgs fields \cite{Blum:2016szr}. The operators listed here extend leptoquark EFT operators further. We can see that there are some objects in Table III with higher electric charges. The highest SU(2) charges of accessible states are the SU(2) quadruplet leptoquarks, with quantum numbers ($\overline{3},4,-1/6$) and ($\overline{3},4,-7/6$).  These each contain four differently charged states in the multiplet. The first state has electric charges ($4/3,1/3,-2/3,-5/3$), while the second has charges ($1/3, -2/3, -5/3,-8/3$). 

Additionally, there is a SU(2) LEX doublet in the ($\overline{3},2,-13/6$) representation. This multiplet contains two charged states: ($-5/3,-8/3$). There are also LEX SU(2) triplets in the ($\overline{3},3,4/3$) and ($\overline{3},3,-2/3$) representations. These multiplets contain the electrically charged components ($7/3,4/3,1/3$) and ($1/3,-2/3,-5/3$), respectively. The higher charge states may be accessible through cascade decays if the masses in the multiplets are split.  

For scalar LEX states, we may split multiplet masses with renormalizable couplings to the Higgs fields. While doublet and triplet splitting operators are discussed later in this work, we might take time in a later work to study the quadruplet mass splitting. There are a few mass splitting possible  contractions with the Higgs fields, for example
\begin{equation}
\lambda_4(H_i\phi^i_{jk})(H^{\dagger i}\phi_i^{\dagger jk})+\lambda^{'}_4(H_l\phi_{ijk})(H^{\dagger l}\phi^{\dagger ijk}) \ .
\end{equation}
These will be mass splitting of order $\sqrt{\lambda_4}v_h$. For TeV scale quadruplet masses, it is reasonable to assume bounds on the Higgs coupling of order 0.1 due to Higgs constraints. Therefore, one may expect mass splittings of the multiplet in the 10's of GeV range.

\subsection{$\boldsymbol{DDLQ\phi}$}
The last operator type involving a LEX scalar that we need to consider contains two derivatives, two SM fermions (one quark and one lepton) and the LEX scalar state. Operators of  this structure were studied in a previous work on the lepton-quark portal.  There are three different Lorentz structures that could be implemented with the two-derivative feature:
\begin{align}
    D^\mu D_\mu \quad , \quad D_\mu D_\nu \gamma^\mu \gamma^\nu \quad , \quad D_\mu D_\nu \sigma^{\mu\nu} \ .
\end{align}
As shown in previous work, the  derivative structures involving two derivatives, $D^\mu D^\nu \sigma_{\mu\nu}$, contain six operators and three integration by parts constraints. The remaining three operators, which we choose to be $D^\mu D^\nu$  acting on a single field, may be reduced with field theory identities to operators of the type in Section 2.1. 

We now consider the $D^\mu D_\mu$ structure. For this structure and the three fields, there are six different ways that the derivatives could be arranged. However, the option where both derivatives act on the lepton do not contribute to the lepton-gluon portal. There are then 3 integration by parts relations involving these operators, leaving us with only two unique operators. We choose to keep the operators
\begin{align}
   \overline{L^c} D^2 Q  \phi \quad ,  \quad \overline{L^c} D_\mu Q D^\mu \phi   .
\end{align}

In the first operator, we invoke the relation
\begin{equation}
D^2f=\slashed{D}\slashed{D}f -i\sigma^{\mu\nu}F_{\mu\nu}f \ .
\end{equation}
Plugging in the first term of Eq. 11, we see a redundancy with other two-derivative operators of the form $D_{\mu}D_{\nu}\gamma^{\mu}\gamma^{\nu}$. Meanwhile, plugging in the second term results in operators of the form of those in Table I. 

We now turn our attention to the second term in Eq. 10 \cite{Lehman:2014jma}.
\begin{align}
\begin{split}
\overline{L^c} D_\mu Q D^\mu \phi &=\frac{1}{2}\overline{L^c}D^{\nu}(\gamma^{\mu}\gamma^{\nu}+\gamma^{\nu}\gamma^{\mu})QD_{\mu}\phi \\
&=\frac{1}{2}\overline{L^c}(\gamma^{\mu}\slashed{D}+\slashed{D}\gamma^{\mu})QD_{\mu}\phi
\end{split}
\end{align}
These are operators have the form $D_{\mu}D_{\nu}\gamma^{\mu}\gamma^{\nu}$. We have thus removed all of the $D^{\mu}D_{\mu}$, or shown that they are equivalent to operators of other categories.

Finally, we examine the last two-derivative Lorentz structure. There are 12 total operators of type $D_{\mu}D_{\nu}\gamma^{\mu}\gamma^{\nu}$. There are six different ways to arrange the two derivatives among three fields (rearrangement of derivatives gives no new terms). Then, the fermion bilinear may have gamma's in two orderings. However, there are six integration by parts relations. With these, we can remove operators with  gamma ordering $\gamma^{\nu}\gamma^{\mu}$. As we will show, the remaining six operators vanish from the portal or are redundant with previous operators.

There are 3 operators of the type $\slashed{D}\slashed{D}$. These are
\begin{equation}
\slashed{D}\slashed{D}\overline{L^c}Q\phi, \qquad \overline{L^c}\slashed{D}\slashed{D}Q\phi, \qquad
\slashed{D}\overline{L^c}\slashed{D}Q\phi \ .
\end{equation}
The operator with two $\slashed{D}$'s acting on the lepton does not contribute to the portal. In the second and third operators, we can invoke a fermion equation of motion 
\begin{equation}
\slashed{D}Q=y_uHq_u+y_dH^{\dagger}q_d \ .
\end{equation}
This reduces these operators to those of the form $DLQ\phi H$ in Section 2.3. 
The remaining three operators are
\begin{equation}
\overline{L^c}\gamma^{\mu}\gamma^{\nu}QD_{\mu}D_{\nu}\phi=
i/2 \overline{L^c}\sigma^{\mu\nu}QF_{\mu\nu}\phi \ ,
\end{equation}
and
\begin{equation}
\overline{L^c}\gamma^{\mu}\slashed{D}QD_{\mu}\phi, ~~~ \overline{L^c}\overleftarrow{\slashed{D}}\gamma^{\nu}QD_\nu\phi
\end{equation}
can use fermion equations of motion for $\slashed{D}f$ to reduce to an operators of the form $DLQ\phi$. 

Of the six operators with gamma ordering $\gamma^{\nu}\gamma^{\mu}$ three
\begin{equation}
D^{\mu}D^{\nu}\overline{L^c}\gamma^{\nu}\gamma^{\mu}Q\phi, ~\overline{L^c}\gamma^{\nu}\gamma^{\mu}QD^{\mu}D^{\nu}\phi, ~~D^{\nu}\overline{L^c}\gamma^{\nu}\gamma^{\mu}D^{\mu}Q\phi
\end{equation}
may be immediately removed or are redundant. The first does not contribute to the lepton gluon portal. The next can be reduced to the form $GLf\phi$ using the covariant derivative identity. The last may be reduced using fermion equations of motion to an operator of form $DLQ\phi$. The remaining operators may be removed by integration by parts identities.
\begin{align}
D^{\mu}D^{\nu}\overline{L^c}\gamma^{\nu}\gamma^{\mu}Q\phi=
D^{\nu}\overline{L^c}D^{\mu}\gamma^{\nu}\gamma^{\mu}Q\phi+
D^{\nu}\overline{L^c}\gamma^{\nu}\gamma^{\mu}QD^{\mu}\phi \\
\overline{L^c}D^{\mu}D^{\nu}\gamma^{\nu}\gamma^{\mu}Q\phi=D^{\mu}\overline{L^c}D^{\nu}\gamma^{\nu}\gamma^{\mu}Q\phi+\overline{L^c}D^{\nu}\gamma^{\nu}\gamma^{\mu}QD^{\mu}\phi \\
\overline{L^c}\gamma^{\nu}\gamma^{\mu}QD^{\mu}D^{\nu}\phi=
D^{\mu}\overline{L^c}\gamma^{\nu}\gamma^{\mu}QD^{\nu}\phi+
\overline{L^c}D^{\mu}\gamma^{\nu}\gamma^{\mu}QD^{\nu}\phi\\
D^{\mu}\overline{L^c}D^{\nu}\gamma^{\nu}\gamma^{\mu}Q\phi=\overline{L^c}D^{\mu}D^{\nu}\gamma^{\nu}\gamma^{\mu}Q\phi+\overline{L^c}D^{\nu}\gamma^{\nu}\gamma^{\mu}QD^{\mu}\phi\\
D^{\mu}\overline{L^c}\gamma^{\nu}\gamma^{\mu}QD^{\nu}\phi=\overline{L^c}D^{\mu}\gamma^{\nu}\gamma^{\mu}QD^{\nu}\phi+\overline{L^c}\gamma^{\nu}\gamma^{\mu}QD^{\mu}D^{\nu}\phi\\
\overline{L^c}D^{\mu}\gamma^{\nu}\gamma^{\mu}QD^{\nu}\phi=D^{\mu}\overline{L^c}\gamma^{\nu}\gamma^{\mu}QD^{\nu}\phi+\overline{L^c}\gamma^{\nu}\gamma^{\mu}QD^{\mu}D^{\nu}\phi
\end{align}
with the results about Eq. 15 and using fermion equations of motion in the third fourth and fifth lines we find that operators of the two derivative type are redundant or do not contribute to the portal.

These are also redundant with previously discussed structures. There are no new operators of this type.

Looking at the entire list of scalar operators, those in Tables II and III have baryon and lepton number consistent with that of leptoquarks. The operators in Table I have the most interesting combinations of lepton and baryon number; while there are some operators consistent with the discreet charges of leptoquarks, we also find operators with LEX states of double lepton number.

\section{Spin 1/2 LEX States}
We now shift our focus from operators involving scalar LEX states to those that are spin-1/2 (fermions). In these operators, we again require a SM lepton, in addition to the LEX fermion. Adding two additional fermions to the operator would bring us to dimension 6, with no gluon coupling. As such, in this paper where we enumerate operators of dimension 6 or less, we only allow operators with two fermions. The lack of a quark in our operators means that there must be an SU(3) field strength tensor in order to generate the gluon coupling. Additional Higgs fields or derivatives may be allowed.
\subsection{$\boldsymbol{GL\psi}$}
We start by considering the easiest possible operators in the $\ell$-G portal which involve fermionic LEX states. These operators involve only one lepton, one LEX fermion, and the SU(3) field strength tensor. Operators of this type, which are dimension 5, are shown in Table IV.
\begin{center}
\begin{tabular}{|c|c|c|c|}
\hline 
Operators with fermion LEX fields & (SU(3),SU(2),Y) & B & L \tabularnewline
\hline 
\hline 
$ G_{\mu\nu} \overline{\ell} \sigma^{\mu\nu} \psi$ & ($8,1,-1$) & 0 & 1 \tabularnewline
\hline 
$ G_{\mu\nu} \overline{L}^i \sigma^{\mu\nu} \psi_i$ & ($8,2,-1/2$) & 0 & 1 \tabularnewline
\hline

\end{tabular}
\\
Table IV. Operators containing $GL\psi$
\end{center}
These two dimension 5 fermionic operators constitute the normal lepto-gluons which appear in the phenomenological literature. In both operators, the LEX state is a color octet fermion with color indices contracted with the SU(3) field strength tensor. These operators contain a single SM lepton, and we may assign the LEX state lepton number 1 in lepton-number conserving theories. 

There are two possible SU(2) charges of the LEX state. In the case that the operator contains a left-handed lepton, the LEX state is an electroweak doublet, while in the case of a right-handed lepton the LEX state is an electroweak singlet. The LEX state in the first operator of Table IV contains a single octet of electric charge $-1$.  The second operator contains a fermionic color octet doublet with charges $(0,-1)$ and will couple to a gluon-neutrino pair or gluon-charged lepton pair, respectively.

\subsection{$\boldsymbol{GL\psi H}$}
Next, we consider dimension 6 operators with a fermionic LEX state. These operators contain a lepton, LEX fermion, SU(3) field strength tensor, and Higgs field. These dimension 6 operators are displayed in Table V. 
\begin{center}
\begin{tabular}{|c|c|c|c|}
\hline 
Operators with fermion LEX fields & (SU(3),SU(2),Y) & B & L \tabularnewline
\hline 
\hline 
$ G_{\mu\nu} \overline{\ell} \sigma^{\mu\nu} \psi^i H_i$ & ($8,2,3/2$) & $0$ & $1$ \tabularnewline
\hline 
$ G_{\mu\nu} \overline{\ell} \sigma^{\mu\nu} \psi_i H^{\dagger i}$ & ($8,2,-1/2$) & $0$ & $1$ \tabularnewline
\hline
\hline
$ G_{\mu\nu} \overline{L}^i \sigma^{\mu\nu} \psi_i^j H_j$ & ($8,3,-1$) & $0$ & $1$ \tabularnewline
\hline 
$ G_{\mu\nu} \overline{L}^i \sigma^{\mu\nu} \psi H_i$ & ($8,1,-1$) & $0$ & $1$ \tabularnewline
\hline 
 
$ G_{\mu\nu} \overline{L}^i \sigma^{\mu\nu} \psi_{ij} H^{\dagger j}$ & ($8,3,0$) & $0$ & $1$ \tabularnewline
\hline 
$ G_{\mu\nu} \overline{L}^i \sigma^{\mu\nu} \psi H^{\dagger}_i$ & ($8,1,0$) & $0$ & $1$ \tabularnewline
\hline 

\end{tabular}
\\
Table V. Dimension 6 Operators containing $GL\psi H$. LEX states are color octet fermions.
\end{center}
 Once again, the LEX state must be a color octet due to the SU(3) field strength tensor. The SU(2) charge of the LEX state again depends on the handedness of the lepton. In the case of a right-handed lepton, the only SM field with non-singlet SU(2) charge is the Higgs. The LEX state is thus also an SU(2) doublet. In the case of the left-handed lepton, we again follow the SU(2) product rule $2\otimes2 = 1\oplus 3$, and the LEX field is an electroweak singlet or triplet. Again, we can assign the LEX state lepton number 1.

 We note that operators may contain either an $H$ or $H^{\dagger}$, which will have interesting consequences for the electric charges or accessible LEX states. Upon insertion of a Higgs vev, these operators contain terms that are of effective dimension 5 with an effective cutoff of $v_h/\Lambda$. 

The first two operators in Table V involve a right-handed, charged lepton. In these operators, the LEX states are SU(2) doublets with SU(2) indices contacted with a Higgs field. 
In the first operator of Table V, we see that the LEX state has hypercharge $3/2$. As such, the multiplet contains both a singly charged state and a doubly charged state, $(\psi_{-}, \psi_{--})$. The doubly charged state may be obtained through electroweak cascade decay of the $\psi_{-}$ state if the masses in the multiplet are split.
The second operator involves a Higgs conjugate. Therefore, the accessible LEX octet multiplet has electric charges $(\psi_0,\psi_{-})$. For both of the top two operators, once a Higgs vev is inserted we recover an effective dimension 5 operator which couples the charge $-1$ fermion component to the gluon and lepton: $\frac{v_h}{\Lambda^2} G_{\mu\nu}\overline{\ell}\sigma^{\mu\nu}\psi_{-}$.

The next four operators involve LEX states that are SU(2) triplets or singlets. In these operators, a left-handed SM lepton appears in the operators. We will start our discussion of these operators by looking at the ones containing color octet, electroweak triplet LEX states. 

We find a LEX fermion in the ($8,3,0$) representation in a dimension 6 operator that involves an insertion of $H^{\dagger}$. This field is the strong-weak bi-adjoint.  The authors have previously made forays into the phenomenology of bi-adjoint scalars. The bi-adjoint has three charged states in the multiplet: $(\psi_{-}, \psi_0, \psi_{+})$. If instead we consider the dimension 6 operator with an $H$ insertion, we find a LEX state in the  ($8,3,-1$) representation. This is a U(1)-charged bi-adjoint whose multiplet contains a doubly charged state. Written out, the multiplet contains $(\psi_{--}, \psi_{-}, \psi_{0})$.  Upon an insertion of the Higgs vev, we get   $\frac{v_h}{\Lambda^2} G_{\mu\nu}\overline{\ell}\sigma^{\mu\nu}\psi_{-}$ and  $\frac{v_h}{\Lambda^2} G_{\mu\nu}\overline{\nu}\sigma^{\mu\nu}\psi_{0}$.  

Finally, color octet, SU(2) singlet fermions are accessible through the lepton-gluon portal. The particle in the ($8,1,-1$) representation has the same charges as the standard lepto-gluon which also appears in Table IV. A pure color octet in the ($8,1,0$) representation is also accessible. However, this fermion still has lepton number 1. This field has contains only a $\psi_0$ state. Once Higgs vevs are inserted, this yields an effective dimension 5 operator $\frac{v_h}{\Lambda^2} G_{\mu\nu}\overline{\nu}\sigma^{\mu\nu}\psi_{0}$. This operator couples the neutral fermion to a neutrino. 

We note that for LEX states in the doublet or triplet representations of SU(2), the charge components may be split through the dimension 5 effective operators 
\begin{equation}
\frac{1}{\Lambda}H^i\overline{\psi_i}H^{\dagger j}\psi_j \quad \mathrm{and} \quad \frac{1}{\Lambda}H^i\overline{\psi_{ij}}H^{\dagger k}\psi_{jk}
\end{equation}
if only one species of LEX particle appears in the theory. These mass splittings will generally be small, of order $v_h^2/\Lambda$. For cut-offs in the few TeV range, this will lead to GeV-order mass splittings.

\subsection{$\boldsymbol{DGL\psi}$}

Finally, we consider operators that include a SM lepton, LEX fermion, SU(3) field strength tensor, and one derivative. The lepton in question may be right-handed or left-handed.
We have 9 initial types of operators:
\begin{align}
\begin{split}
G_{\mu\nu}\slashed{D} \overline{\ell}\sigma^{\mu\nu}\psi, \qquad & ~~G_{\mu\nu}\overline{\ell}\sigma^{\mu\nu}\slashed{D}\psi,  \qquad  \qquad D^{\rho}G_{\mu\nu}\overline{\ell}\gamma^{\rho}\sigma^{\mu\nu}\psi,
\\
G_{\mu\nu}D^{\mu}\overline{\ell}\gamma^{\nu}\psi, \qquad & ~D^{\mu}G_{\mu\nu}\overline{\ell}\gamma^{\nu}\psi, \qquad \qquad~G_{\mu\nu}\overline{\ell}\gamma^{\nu}D^{\mu}\psi,
\\
D^{\rho}G_{\mu\nu}\overline{\ell}\gamma^{\mu}\gamma^{\rho}\gamma^{\nu}\psi, \qquad  &G_{\mu\nu}D^{\rho}\overline{\ell}\gamma^{\mu}\gamma^{\rho}\gamma^{\nu}\psi, \qquad  ~~~G_{\mu\nu}\overline{\ell}\gamma^{\mu}\gamma^{\rho}\gamma^{\nu}D^{\rho}\psi \ .
\end{split}
\end{align}
The last three operators may be reduced to the sum of the first two operator types by applying the anticommutation rule
\begin{equation}
\eta^{\rho \mu}=\frac{1}{2}(\gamma^{\rho}\gamma^{\mu}+\gamma^{\mu}\gamma^{\rho}) \ .
\end{equation}
For example,
\begin{align}
\begin{split}
G_{\mu\nu}\overline{\ell}\gamma^{\mu}\gamma^{\rho}\gamma^{\nu}D^{\rho}\psi &= -G_{\mu\nu}\overline{\ell}\gamma^{\mu}\gamma^{\nu}\gamma^{\rho}D^{\rho}\psi+G_{\mu\nu}\overline{\ell}\gamma^{\mu}g^{\rho\nu}D^{\rho}\psi
\\
&= -G_{\mu\nu}\overline{\ell}\sigma^{\mu\nu}\slashed{D}\psi+G_{\mu\nu}\overline{\ell}\gamma^{\mu}D^{\nu}\psi \ ,
\end{split}
\end{align}
where in the last line we have used $\gamma^{\mu}\gamma^{\nu}= 1/2g^{\mu\nu}+1/2 \sigma^{\mu\nu}$.

In the first two operators we may use the fermion equations of motion,
\begin{equation}
\slashed{D}f=yHf
\end{equation}
or \begin{equation}
\slashed{D}f=yH^{\dagger}f \ .
\end{equation}
As such, we see that the first two types of operators reduce to those of the form $GL\psi$H and are thus redundant with those in Table V. Invoking the integration by parts relation 
\begin{equation}
-D^{\rho}G_{\mu\nu}\overline{\ell}\gamma^{\rho}\sigma^{\mu\nu}\psi=G_{\mu\nu}\slashed{D} \overline{\ell}\sigma^{\mu\nu}\psi+G_{\mu\nu}\overline{\ell}\sigma^{\mu\nu}\slashed{D}\psi \ ,
\end{equation}
we may eliminate all three operators of the first type.

Now we consider the next three operators, where in each the derivative $D_{\mu}$ acts on a different field. However, the three operators are connected by one integration by parts relation. 
\begin{equation}
G_{\mu\nu}D^{\mu}\overline{\ell}\gamma^{\nu}\psi = D^{\mu}G_{\mu\nu}\overline{\ell}\gamma^{\nu}\psi+G_{\mu\nu}\overline{\ell}\gamma^{\nu}D^{\mu}\psi
\end{equation}
We note that the operator with $D^{\mu}G_{\mu\nu}$ is proportional to an equation of motion. 
\begin{equation}
 D^{\mu}G_{\mu\nu}=\Sigma\overline{Q}\gamma^{\nu}Q+\Sigma\overline{q}\gamma^{\nu}q + \Sigma\overline{\psi}_{LEX}\gamma^{\nu}\psi_{LEX}
\end{equation} 
This operator will give 0 amplitude for any on-shell gluon in the process. However, the operator can still be considered to contribute to processes involving an off-shell gluon. We can choose, however to eliminate the operator proportional to  $D^{\mu}G_{\mu\nu}$ as redundant by invoking the  constraint in Eq. 31. This leaves us with the operators
\begin{equation}
G_{\mu\nu}D^{\mu}\overline{\ell}\gamma^{\nu}\psi \quad \mathrm{and} \quad G_{\mu\nu}\overline{\ell}\gamma^{\nu}D^{\mu}\psi \ .
\end{equation}
These two equations cannot be further reduced. For example, taking one operator, we find
\begin{equation}
G_{\mu\nu}D^{\mu}\overline{\ell}\gamma^{\nu}\psi = G_{\mu\nu}D^{\rho}\eta_{\rho \mu}\overline{\ell}\gamma^{\nu}\psi \ .
\end{equation}
Using Eq. 26, we transform the operator above to
\begin{equation}
\frac{1}{2}G_{\mu\nu}\overline{\ell}[\overleftarrow{\slashed{D}} \gamma^{\mu}+\gamma^{\mu}\overleftarrow{\slashed{D}}]\gamma^{\nu}\psi \ .
\end{equation}
The first term may be eliminated as redundant using the fermion equation of motion as before. The second term can be integrated by parts.
\begin{equation}
\frac{1}{2}D_{\rho}G_{\mu\nu}\overline{\ell}\gamma^{\mu}\gamma^{\rho}\gamma^{\nu}\psi+\frac{1}{2}G_{\mu\nu}\overline{\ell}\gamma^{\mu}\gamma^{\rho}\gamma^{\nu}D_{\rho}\psi \
\end{equation}
As such, we get 
\begin{equation}
=-\frac{1}{2}G_{\mu\nu}D_{\rho}\overline{\ell}\gamma^{\mu}\gamma^{\rho}\gamma^{\nu}\psi
\end{equation}
as the contribution to our portal. Through a relation like Eq. 26, this returns us to our original operator.

We are thus left with two operators in the portal, 
\begin{equation*}
G_{\mu\nu}D^{\mu}\overline{\ell}\gamma^{\nu}\psi \quad \mathrm{and} \quad G_{\mu\nu}\overline{\ell}\gamma^{\nu}D^{\mu}\psi \ .
\end{equation*}
The quantum numbers of the LEX states are the same as those of the standard leptogluons. If the lepton is right-handed then the LEX state will be in the ($8,1,1$) representation. If the lepton is left-handed, the LEX state will be in the ($8,2,-1/2$) representation.

\begin{center}
\begin{tabular}{|c|c|c|c|}
\hline 
Operators with fermion LEX fields & (SU(3),SU(2),Y) & B & L \tabularnewline
\hline 
\hline 
$ G_{\mu\nu}D^{\mu}\overline{\ell}\gamma^{\nu}\psi,~G_{\mu\nu}\overline{\ell}\gamma^{\nu}D^{\mu}\psi $ & ($8,1,-1$) & 0 & 1 \tabularnewline
\hline 
$  G_{\mu\nu}D^{\mu}\overline{L}\gamma^{\nu}\psi,~G_{\mu\nu}\overline{L}\gamma^{\nu}D^{\mu}\psi $ & ($8,2,-1/2$) & 0 & 1 \tabularnewline
\hline 
\end{tabular}
\\
Table VI. Operators containing $DGL\psi$
\end{center}

\section{Dimension 7 Operators}

Because the number of dimension 6 operators is so limited, we briefly mention some interesting features of dimension 7 operators. 

\subsection*{Operators of type $GFL\psi$}

Among the most striking dimension 7 operators are a type that contain two SM gauge field strength tensors, a SM lepton, and a fermionic LEX state. In the $\ell$-G portal, one of these field strength tensors must be a $G_{\mu\nu}$. The remaining field strength tensor may be $G_{\mu\nu}$, $W_{\mu\nu}$ (the SU(2) field strength tensor), or $B_{\mu\nu}$ (the U(1) field strength tensor). Operators of this type allow the most spectacular charge assignments for the LEX fermion, as we will see.

\begin{tabular}{|c|c|c|c|}
\hline 
 \multicolumn{4}{|c|}{Dimension 7 Operators} \\
\hline 
Operator with scalar LEX field & (SU(3),SU(2),Y) & B & L\tabularnewline
\hline 
\hline 
\multirow{2}{*}{$G_{\mu\nu}G^{\mu\nu}\overline{\ell}\psi$} & ($1,1,-1$), ($8,1,-1$), ($10,1,-1$), & \multirow{2}{*}{0} & \multirow{2}{*}{1}\tabularnewline
& ($\overline{10},1,-1$), ($27,1,-1$) & & \tabularnewline
\hline 
\multirow{2}{*}{$G_{\mu\nu}G^{\mu\nu}\overline{L^{i}}\psi_{i}$} & ($1,2,-1/2$), ($8,2,-1/2$), ($10,2,-1/2$), & \multirow{2}{*}{0} & \multirow{2}{*}{1} \tabularnewline
& ($\overline{10},2,-1/2$)($27,2,-1/2$) & & \tabularnewline
\hline 
$W_{\mu\nu j}^{i}G^{\mu\nu}\overline{\ell}\psi_{i}^{j}$ & ($8,3,-1$) & 0 & 1\tabularnewline
\hline 
$W_{\mu\nu}^{ij}G^{\mu\nu}\overline{L^{k}}\psi_{ijk}$ & ($8,4,-1$) & 0 & 1\tabularnewline
\hline 
$W_{\mu\nu j}^{i}G^{\mu\nu}\overline{L^{j}}\psi_{i}$ & ($8,2,-1$) & 0 & 1\tabularnewline
\hline 
$B_{\mu\nu}G^{\mu\nu}\overline{\ell}\psi$ & ($8,1,-1$) & 0 & 1\tabularnewline
\hline 
$B_{\mu\nu}G^{\mu\nu}\overline{L^{i}}\psi_{i}$ & ($8,2,-1/2$) & 0 & 1\tabularnewline
\hline 
\end{tabular}
\begin{center}
Table VII. Operators of type $GFL\psi$
\end{center}

In the operators shown in Table VII, Lorentz indices are contracted between the two field strength tensors which couple to the fermionic scalar current (unlike operators of the type $GL\psi$ and $GL\psi H$ which couple to the fermion tensor current). We see that the operator that contains two SU(3) field strength tensors allows the most interesting color charge charge assignments for the LEX field. To make this charge assignment, we have used the SU(3) tensor product rule
\begin{equation}
8\otimes8=1\oplus8\oplus8\oplus10\oplus\overline{10}\oplus27 \ .
\end{equation}
The 27 is an SU(3) object with two upper and two lower SU(3) fundamental indices. The 10 has three upper and zero lower indices, while the $\overline{10}$ has three lower and zero upper indices\cite{Coleman:1965afp}.  The operators that contain two strong field strength tensors can contain either a left- or right-handed SM lepton. LEX states are assigned doublet SU(2) charge for operators containing a left-handed SM fermion, and SU(2) singlet charge for operators containing right-handed SM fermions. 

The next set of operators may contain one SU(3) field strength tensor and one SU(2) field strength tensor. In these operators, the LEX state must be a color octet. For operators containing a SM fermion that is an SU(2) singlet, the LEX state must be an SU(2) adjoint in order to contract SU(2) indices with the $W_{\mu}{\nu}$.  For operators containing an left-handed SM fermion, we may invoke the SU(2) tensor product rule
\begin{equation}
2\otimes 3 = 4\oplus2
\end{equation}
which gives us SU(2) doublet and quadruplet options for the LEX state. The accessible LEX states are thus exotic fermionic octets with higher SU(2) representations. These operators are phenomenologically interesting because they couple a lepton-gluon pair to an electroweak gauge boson and an exotic LEX fermion.

Finally, this class of dimension 7 operators may involve a U(1) field strength tensor. In this case, the LEX state must be a color octet. If the operator involves a right-handed SM lepton, the LEX state will be an SU(2) singlet; if it involves a left-handed SM lepton, the LEX state will be an SU(2) doublet. These LEX fermions have the same charge assignment as standard lepto-gluons.  The operators will couple a lepton-gluon pair to a LEX fermion and a photon or Z boson.  Interestingly, the final operator in the list  allows the coupling of a neutral LEX fermion to a neutrino, gluon, and photon or Z boson.

We note that there is also a class of operators with alternate Lorentz structure. These operators have a pseudo-scalar fermion current and one dual field strength tensor. The operators have the same field content as each of the operators in Table VII and are of the form
\begin{equation}
\tilde{F}_{\mu\nu}^1 F^{ 2 \mu\nu}\overline{f}\gamma^5\psi \ .
\end{equation}

The first operator allows LHC gluon-gluon fusion processes to a SM lepton and higher-color multiplet. Operators with SU(2) or U(1) field strength tensors allow 2-3 LHC production processes that involve the production of a lepton and exotic octet with an associated hard photon, W, or Z boson from gluon fusion: $gg\rightarrow W/Z/\gamma+\psi+l$.

These operators also allow 2-3 processes in lepton colliders which produce exotic octets of higher SU(2) charge. These processes exchange a t-channel electroweak boson and produce a forward lepton, a gluon, and a LEX fermion: $l^+ l^- \rightarrow g+l+\psi$.

At dimension 7, an additional Higgs may be added to any of the operators in Tables I-V.  This will make higher SU(2)-charged LEX states available. For example, the exotic scalar and fermionic octets in Tables I and V may be extended to include SU(2) quadruplets. Dimension 7 operators also allow for the possibility of four-fermion interactions with an additional covariant derivative. These would be found in operators of type $Dfff\psi$. Within these four-fermion interactions, we can have LEX states with either double lepton number or double baryon number.  Specifically, those operators with two leptons ($DLLQ\psi$) have double lepton number and non-zero baryon number. Similarly, operators of type $DLQQ\psi$ would provide access to states with double baryon number,and unit lepton number. In the case that all of the SM fermions are left-handed, the LEX state could also have SU(2) charge up to that of a quadruplet.

\section{LHC production of Exotic Octets }

Here we present LHC production cross sections for some of the exotic fermionic octet models presented in this work. Among these are the extended fermionic lepto-gluons found in Table V.
 
 Specifically, we are interested in processes where an initial state gluon or gluons create a lepton-LEX state pair, sometimes in conjunction with a Higgs boson. Lepton associated production is a known hadron collider production mechanism for standard lepto-gluons. Here, we compare the production cross-sections of our menagerie of six types of exotic fermionic octet multiplets---two with standard lepto-gluon charge assignments and four with more exotic charge assignments. Depending on the model in question, the final state lepton may be either the charged lepton or a neutrino produced in association with unit charges or neutral component of the exotic LEX multiplet. We once again note that operators in Table V are dimension 6.  In Fig. 3, we show LEX-lepton associated production diagrams for the gluon induced processes. 
\begin{center}
\includegraphics[width=0.5\linewidth]{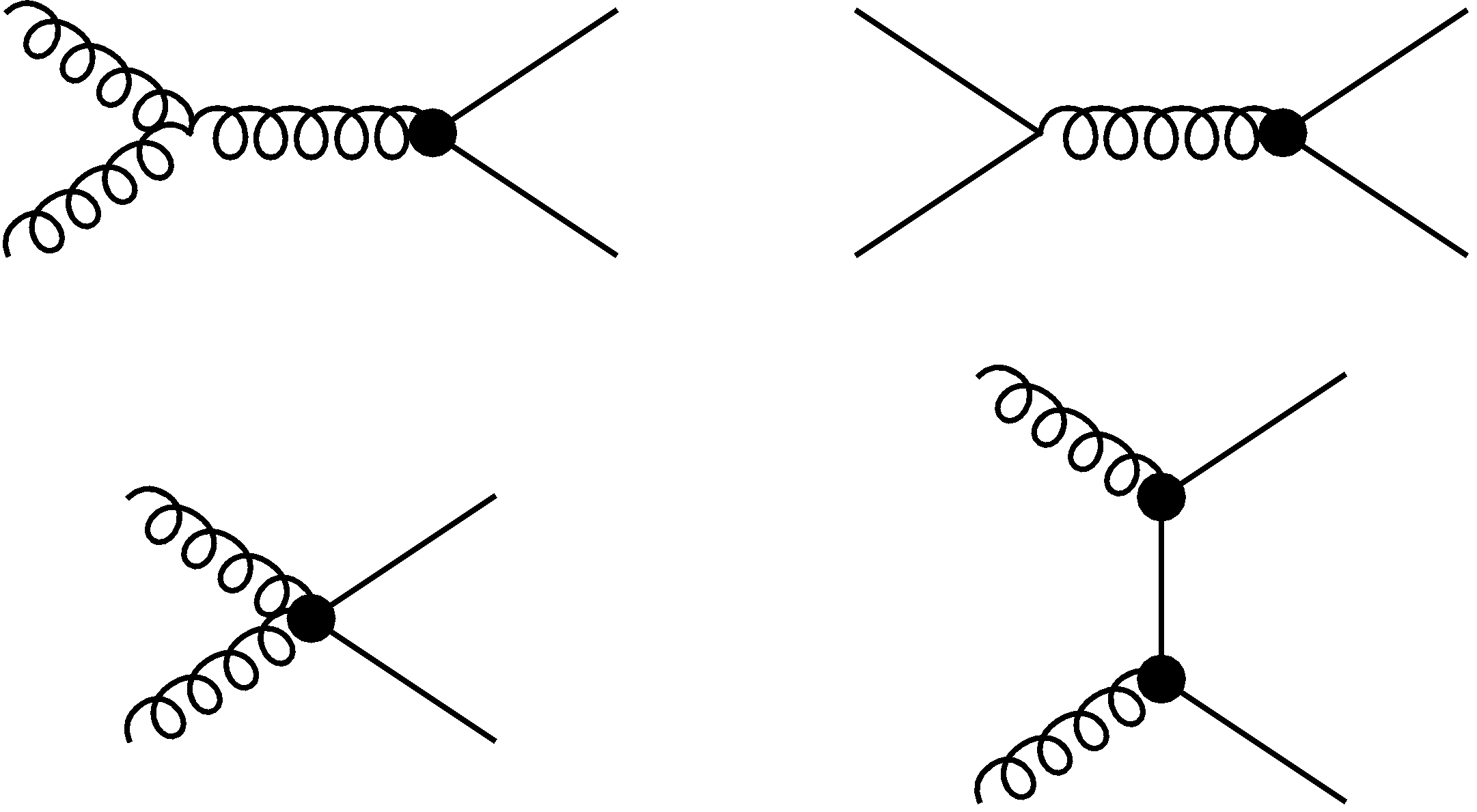}\\
{Figure 3. Feynman diagrams for production of the extended lepto-gluons and associated lepton}
    \label{fig:placeholder}\\
\end{center}

 To compute LHC production processes, we have input the six models of Table V using Feynrules \cite{FR_2,FR_OG} to create models in UFO \cite{UFO} format. Models were fed to Madgraph5@NLO\cite{MG5_EW_NLO} for computation of the production of leading-order cross sections.
 
 In Fig. 4, we display cross sections for the process vs LEX fermion mass for two production modes involving SM anti-leptons, $pp\rightarrow l^+ \psi^-$ and $pp\rightarrow l^+ \psi^- h$. Models in this figure -with associated LEX charge assignments- are

\begin{align}
\begin{split}
(8,2,3/2)\qquad\qquad~~(8,1,-1)~~~~~~~\qquad(8,2,-1/2)\\
  G_{\mu\nu} \overline{\ell} \sigma^{\mu\nu} \psi^i H_i \qquad G_{\mu\nu} \overline{L}^i \sigma^{\mu\nu} \psi H_i \qquad G_{\mu\nu} \overline{\ell} \sigma^{\mu\nu} \psi_i H^{\dagger i}
  \end{split}
\end{align}

In Fig. 4, we see fb-level production cross sections for lepton-LEX associated production for LEX states up to 1.4 TeV. The 2 to 3 production process which produces a LEX state, lepton, and Higgs particles are generally smaller by 1-2 orders of magnitude.

\begin{center}
\includegraphics[width=0.5\linewidth]{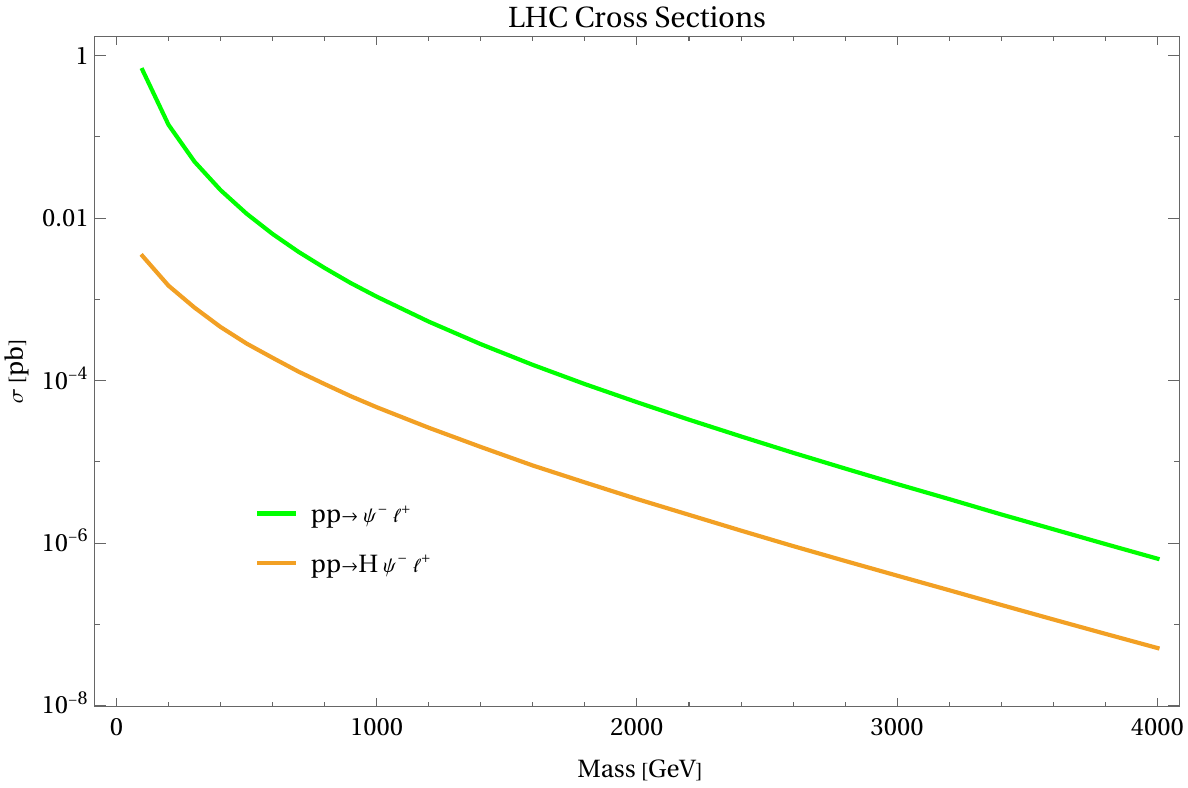}\\
{Figure 4. Results for associated lepton production for the ($8,2,3/2$), ($8,2,-1/2$), and ($8,1,-1$) $\psi_{-}$ LEX state associates with $\ell^+$ here $\Lambda=5$ TeV}
    \label{fig:placeholder}\\
\end{center}

We may also compare lepton-LEX associated production cross section across models as we see in Figs. 5 and 6 below. In Fig. 5, we compare associated production of anti-lepton $\ell^+$ and a $\psi^- $ state in 5 models. 

\begin{center}
\includegraphics[width=0.5\linewidth]{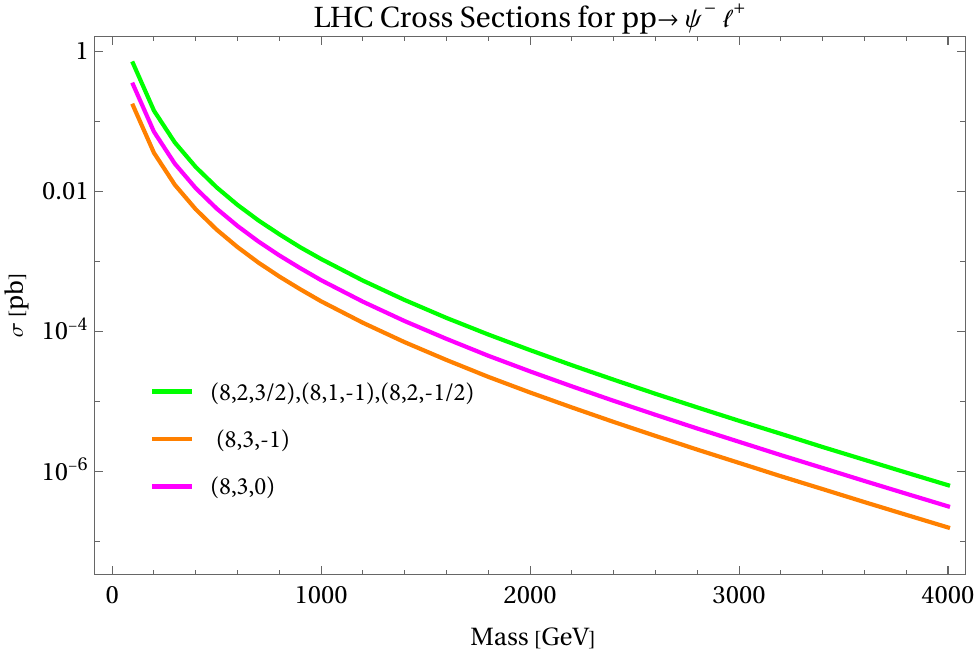}\\
{Figure 5. Results for associated $\ell^+$ $\psi_{-1}$ production for ($8,2,3/2$),($8,2,-1/2$), and ($8,1,-1$)(green); ($8,3,-1$)(orange) and (8,3,0)(pink) Here $\Lambda=5$ TeV}
\end{center}

In the above diagram, the green line corresponds to models given that appear in the figure above, the ($8,2,-1/2$) and ($8,1,-1$) and ($8,2,3/2$) SU(2) doublet and singlet lepto-gluons.  The pink and orange curves correspond to production cross sections for exotic triplet leptoquarks.

\begin{align}
(8,3,-1)\qquad\qquad~~~~(8,3,0)\qquad~~~~\\
  G_{\mu\nu} \overline{L^i} \sigma^{\mu\nu} \psi^j_i H_j \qquad G_{\mu\nu} \overline{L^i} \sigma^{\mu\nu} \psi^j_i H_j^{\dagger}  \qquad 
\end{align}

We note that the sixth LEX state in Table V in the ($8,1,0$) representation does not have a charged component, but only a neutral component. As such, this state does not appear in these production processes. The singlet LEX state in the ($8,1,-1$) representation decays through the same operator. This process is $pp\rightarrow l^+\psi^-\rightarrow l^+l^-g$.  The SU(2) charged multiplets may decay similarly, or may cascade decay to other states in the mutliplet depending on the mass splittings in the theory.

In Fig. 6, we show production processes for the neutral component of the extended lepto-gluons. In these processes, the neutral component is produced in association with a neutrino $pp\rightarrow \nu + \psi_0$.

    \begin{center}
    \includegraphics[width=0.5\linewidth]{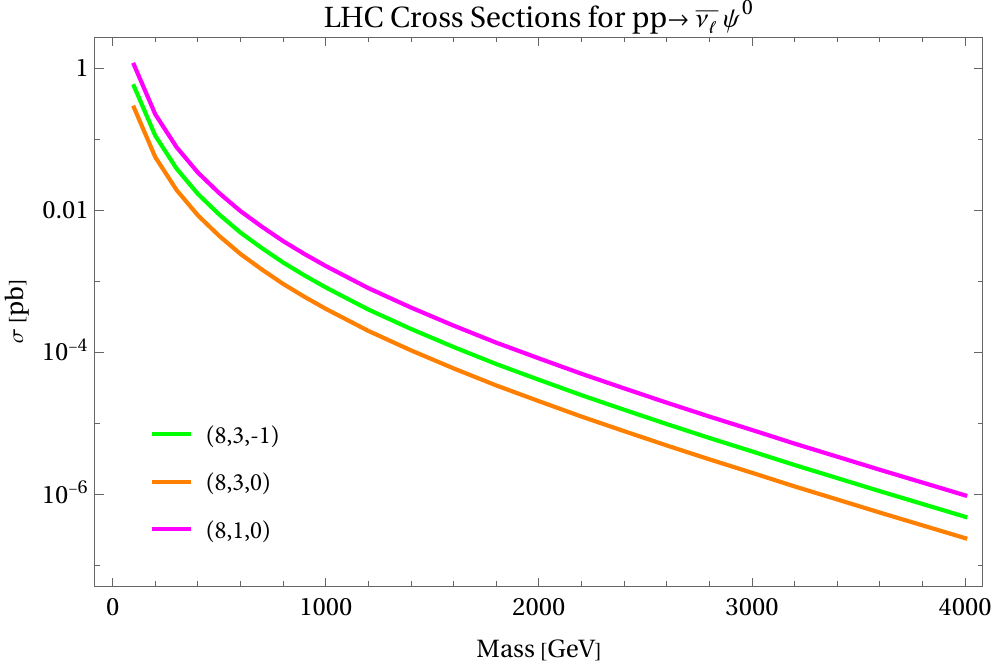}\\
    {Figure 6. Results for associated neutrino production for ($8,3,-1$)(green), ($8,3,0$)(orange), and ($8,1,0$)(pink) with $\psi_{0}$ LEX. Here $\Lambda=5$ TeV}
    \label{fig:placeholder}\\
    \end{center}

 In Fig. 6, the pink line corresponds to the ($8,1,0$) color octet singlet model. This LEX state does not contain any charged components. The other two models are the neutral components of the exotic bi-adjoint lepto-gluons ($8,3,-1$)(green) and ($8,3,0$)(orange).  We see that the neutral component production occurs with a significant missing energy. Production cross section of the neutral octets remain above 1fb for masses is the 1.4 TeV range. The singlet ($8,1,0$) LEX state must decay to a gluon and neutrino, revealing a monojet topology $pp\rightarrow \nu+\psi_0\rightarrow \nu \nu g$.  The ($8,3,-1$) and ($8,3,0$) multiplets may either decay through the same operator, or may cascade decay to the charged components of the multiplet, if the masses are split.

\section{Conclusion}
We have enumerated the operators of the lepton-gluon portal, up to dimension 6. These operators include all of the ways that a light exotic BSM state can couple to the SM through interactions with a lepton-gluon pair. We have enumerated exotic states beyond the standard lepto-gluon. Some of the accessible exotic particles  are in higher dimensional representations of SU(2) and SU(3). The portal includes scalar SU(3) 15-plets and  sextets some of which also have higher SU(2) charge. The portal includes exotic scalar octets that carry electric higher electric charge and/or double lepton number. This includes, for example, the U(1) charged scalar bi-adjoint of SU(2) and SU(3). Also in the portal are extended types of fermionic leptogluons. These include fermionic species of weak-strong bi-adjoints, exotic charge singlet color octet, and exotic fermionic color octet doublet with doubly charged states. These exotic doublets all carry unit lepton number.   

This portal also includes new effective operators for several known types of exotic particle species including scalar leptoquarks, standard lepto-gluons, and the Manohar-Wise color octet. We explored briefly the LHC phenomenology for some of the exotic states in the paper. We computed LHC production cross section for several production channels for exotic color octet states.

There is much more work to be done in the lepton-gluon portal. Recently, several interesting proposals for studying particle production in lepton-gluon/quark portals have surfaced. This includes study of collisions of high energy neutrinos on fixed target nucleons \cite{Bai:2025pef}, and also lepton-quark/gluon collision studied at LHC considering a lepton pdf in the proton \cite{Almeida:2022udp}.  The exotics of the lepton-gluon portal would be interesting to study in this context. Further, dedicated HL-LHC studies for specific the exotic octets models in this paper would be interesting new avenues for new physics. 

Model building in the octet sector would also be very interesting. The authors recently built some simple models containing an ($8,3,0$) scalar bi-adjoint. The octets in this portal are even more exotic, and it may be worthwhile to explore them in a model building context. A complete EFT operator catalog for any of the Light Exotic states in this work would be another interesting avenue of further study.
Finally, the authors plan to continue cataloging exotic new LEX states by phenomenological portal. This work is the third in such a series and we have plans to explore several future portals.

\bibliographystyle{JHEP}

\bibliography{biblq.bib}

\end{document}